# Tetracene ultrathin film growth on silicon

Jens Niederhausen*‡, Rowan W. MacQueen, and Klaus Lips

*Department ASPIN, Helmholtz-Zentrum Berlin für Materialen und Energie GmbH, Berlin, Germany*

Hazem Aldahhak**‡, Wolf Gero Schmidt, and Uwe Gerstmann

*Lehrstuhl für Theoretische Materialphysik, Universität Paderborn, 33095 Paderborn, Germany*

## Abstract

Inorganic-organic interfaces are important for enhancing the power conversion efficiency of silicon-based solar cells through singlet exciton fission (SF). We elucidated the structure of the first monolayers of tetracene (Tc), a SF molecule, on hydrogen-passivated Si(111) [H-Si(111)] and hydrogenated amorphous Si (a-Si:H) by combining near-edge X-ray absorption fine structure (NEXAFS) and X-ray photoelectron spectroscopy (XPS) experiments with density functional theory (DFT) calculations. For samples grown at or below substrate temperatures of 265 K, the resulting ultrathin Tc films are dominated by almost upright-standing molecules. The molecular arrangement is very similar to the Tc bulk phase, with only slightly higher average angle between the conjugated molecular plane normal and the surface normal ($\alpha$) around 77°. Judging from carbon K-edge X-ray absorption spectra, the orientation of the Tc molecules are almost identical when grown on H-Si(111) and a-Si:H substrates as well as for (sub)mono- to several-monolayer coverages. Annealing to room temperature, however, changes the film structure towards a smaller $\alpha$ of about 63°. A detailed DFT-assisted analysis suggests that this structural transition is correlated with a lower packing density and requires a well-chosen amount of thermal energy. Therefore, we attribute the resulting structure to a distinct monolayer configuration that features less inclined, but still well-ordered molecules. The larger overlap with the substrate wavefunctions makes this arrangement attractive for an optimized interfacial electron transfer in SF-assisted silicon solar cells.

## 1. Introduction

Highly efficient (opto)electronic devices often rely on the rational combination of organic and inorganic materials. Conjugated organic molecules (COMs) exhibit efficient light-matter interactions while inorganic materials provide stability and efficient charge transport. The combination of tetracene (Tc) and silicon (Si) has received considerable attention in recent years due to Tc's potential to enhance the power conversion efficiency of Si-based solar cells: [1-3] The Tc molecules can undergo singlet fission (SF), a process which splits one singlet exciton into two triplet excitons [4]. SF can be exploited to produce additional photocurrent from energy which is ordinarily wasted during carrier thermalization in Si [5, 6].

The inorganic-organic interface is where particle and energy exchange between the two dissimilar materials takes place. Converting the tightly bound triplet excitons in Tc layers into current that can be extracted at the electrodes requires efficient exciton dissociation and/or Dexter energy transfer [7] at the Tc-Si interface. These processes proceed via interfacial electron transfer (IET) and therefore rely on the spatial proximity of the molecular and Si frontier electronic states. Since COMs like Tc have highly anisotropic electronic and chemical structures, their orientation at the interface is of particular importance to achieve an interface morphology that favors efficient IET. As an example, the IET-competing luminescence of Tc on an ultrathin alumina $AlO_x$ layer on $Ni_3Al(111)$ is completely quenched if the molecules are planar or are at least only moderately inclined with respect to the surface, while the luminescence persists for larger Tc inclinations [8-10]. This behavior is attributed to the fact that a smaller Tc inclination gives rise to an increased overlap of $\pi$ orbitals (located above and below the planar Tc core) and the wave function tails of the metal states facilitating, thus, an efficient IET from Tc to the substrate.

In order to optimize IET-based devices, it is thus important to establish preparation protocols which lead to stable well-ordered molecular structures with favorable transport properties at the interface. Here, the orientation of the lowest layer plays a particularly important role. For Tc at non-terminated and therefore strongly interacting Si surfaces [11-13] interlayers with flat-lying molecules are reported. However, flat-lying interlayers establish a further, unfavorable intermolecular interface to the upright-standing molecules in the following layers [14]. An orientation of intermediate inclination thus might be more desirable. A more upright orientation is expected for weakly interacting surfaces.

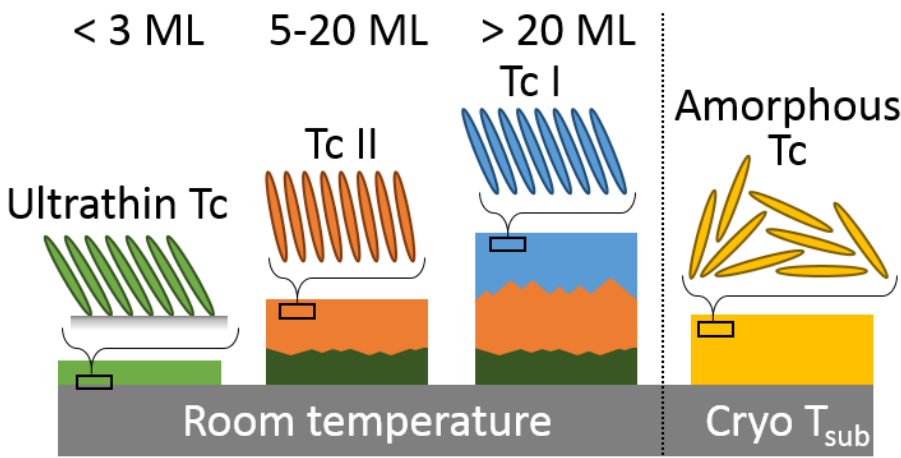

**Figure 1:** Schematic summary of Tc orientations for characteristic Tc films at weakly interacting substrates: Amorphous films form upon deposition at cryogenic $T_{sub}$. At room temperature, the largest well-ordered inclination is found in the so-called thin film phase (Tc II). Lower inclined molecules are found for the bulk phase (Tc I, > 20 ML) as well as for the ultrathin Tc phase (< 3 ML). The darker green indicates a structure transition upon completion of ~ the 3rd Tc ML observed in Ref. [25, 26].

Tc structure formation has been widely studied on several different weakly-interacting surfaces. Thanks to the weak coupling to the substrates, the results are largely transferable. We summarize the current understanding schematically in Fig. 1. The molecular inclination is defined as the average angle between the conjugated molecular plane normal and the surface normal ($\alpha$), as shown in Fig. 2. The smallest $\alpha$ for Tc on weakly-interacting surfaces ($\alpha \sim 43°$) was reported in Ref. 10 for the non-luminescent Tc at AlO$_x$/Ni$_3$Al(111), for which substrate temperatures, $T_{sub}$ < 100 K during film growth was required. However, this immobilized the Tc molecules and frustrated their tendency to self-assemble. This growth mode, thus, does not lead to the formation of ordered islands or crystallites [8-10], but to disordered films that are unfavorable for efficient charge and exciton transport as needed, e.g., in solar cell applications. In addition, films of kinetically frustrated small molecules like Tc usually undergo drastic structural changes upon temperature increase to room temperature (RT). [9, 15, 16]. The use of cryogenic conditions during growth thus appears to be beneficial for decreasing the average Tc inclination angle $\alpha$, but severely limits the applicability to device fabrication.

According to X-ray and low energy electron diffraction (XRD/LEED) spectroscopy, Tc forms polycrystalline films when grown at non-cryogenic temperatures [8, 17-23]. Two ordered phases have been reported (denoted Tc I and Tc II). Both phases feature rather upright-standing molecules, but with considerably different inclination angles $\alpha$ (see Fig. 2). Tc II is often referred as thin film phase since it was found to be predominant in Tc films between approximately 5 and 20-100 monolayers (ML) thick (depending on deposition rate and $T_{sub}$) on $SiO_x$ [17, 18, 20, 21] and H-Si(111) [23]. Based on published crystal data, [20] $\alpha$ within standard thin films (Tc II) is 85.4°. The unit cell of Tc I, in contrast, is measured to be close to that of bulk Tc ($\alpha = 74.5°$ [24]); it is thus called bulk-like and prevails in thicker films above 20-100 ML. The unit cells of Tc I and Tc II each consist of two non-equivalent molecules (see Figure 2). The inclination of these molecules can be also described by the (average) angle between the molecular long axis (LA) and the plane of the underlying surface. This angle is denoted as $\alpha_{LA}$, differs much less for the two non-equivalent molecules than the NEXAFS inclination angle, and thus helps further characterization of the investigated structures (see Fig. 2).

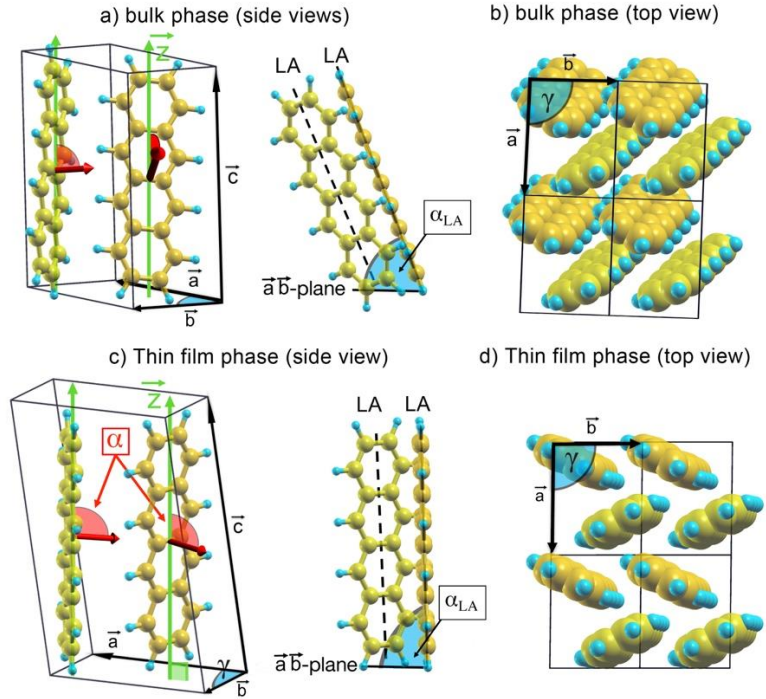

**Figure 2** a)/c) side and b)/d) top-views for the experimentally determined structures of the tetracene bulk (Tc I, *top*) and thin film (Tc II, *bottom*) phases, published in Refs. [24] and [20], respectively. The parameters $\vec{a}$ = 0.781 nm, $\vec{b}$ = 0.603 nm and $\gamma$ = 94.208° ($\vec{a}$ = 0.756 nm, $\vec{b}$ = 0.593 nm and $\gamma$ = 90.10°) define the lateral dimensions of the Tc I (Tc II) phases, whereby the unit cell of each phase includes two non-equivalent molecules. As guide to the eyes, the carbon atoms in the non-equivalent Tc within a unit cell are shown in slightly different colors. It is a unique feature of the bulk-phase that only one molecule is '*simply*' tilted, whereby the second molecule is tilted '*over the edge*', so that one of the two bottom H-atoms is considerably closer to the bottom of the unit cell (assumed to be the $\vec{a}\vec{b}$-plane), see also a). With respect to this bottom $\vec{a}\vec{b}$-plane, the two non-equivalent molecules show almost the same orientation $\alpha_{LA}$ for the long axis (LA), but differently rotated molecular planes. As a consequence, the average NEXAFS angle $\alpha$ between the molecular plane normal (red arrows) and the surface normal ($\vec{z}$-direction, green) deviates from $\alpha_{LA}$ ($\alpha$ is always larger than $\alpha_{LA}$). In the thin film phase, the inclination of the two molecules is at least qualitatively identical. Thereby, Tc II exhibits denser morphology and the Tc molecules are more inclined with respect to the $\vec{a}\vec{b}$-plane ($\alpha_{LA}$ = 85.4° instead of 69.48° in Tc I).

In Tc films consisting of only a few ML, $\alpha$ values of 78° and below, i.e. inclination angles similar to Tc-bulk, have been reported, with a clear trend to much smaller $\alpha$ around 63° for coverages of approximately 1.2 ML or below [10, 11, 25]. Notably, ex-situ atomic force and scanning tunneling microscopy (AFM/STM) revealed that Tc islands with a complete third ML on H-Si(100) [25] and SiO$_2$ [26] show structural characteristics clearly distinct from thinner (1-2 ML) films: They appear to be denser [25] and more stable against interlayer diffusion [26]. Consistently, in situ, real-time grazing incidence XRD found that the crystal structure of Tc films on SiO$_2$ changes within the first few (1-4) MLs [21], suggesting the existence of a proper, but so far not well established ultrathin Tc phase.

The structure of the molecules in Tc films depends not only on the film thickness, but also, and sometimes more decisively, on $T_{sub}$ [9, 22]. Ref. 9 reports structural changes related to changing molecular orientation at around $T_{sub}$ = 210 K; changes around $T_{sub}$ = 250 K and 280 K were interpreted as island formation and partial ordering within the islands, respectively. Significant desorption starts at temperatures of approximately RT [8]. For Tc powder, a phase transition from Tc II to Tc I was reported in the temperature range 275–340 K [27].

In this combined experimental and theoretical work, we study the molecular arrangement at the Tc/H-Si interface for ultrathin films (1 to ~3 ML thickness), for which the reported experimental observations, although decisive for IET, are by far not sufficiently well understood. For this purpose, DFT-assisted near-edge X-ray absorption fine structure (NEXAFS) spectroscopy is applied. By using single-crystalline silicon H-Si(111) and amorphous silicon (a-Si:H) as substrates, we investigate possible template effects [28]. We also investigate the influence of coverage/film thickness. However, the main focus is on the substrate temperature $T_{sub}$ during and after Tc film growth. This parameter controls dewetting, desorption, and phase transitions, and is thus critical for preparing well-defined ultrathin Tc film morphologies.

In the following, we show that (*i*) a hydrogen-terminated Si surface entails weak interaction between Tc and the substrates, (*ii*) the molecular arrangement in ultrathin Tc-films (~1 ML regime) is highly sensitive to the local molecular coverage, and (*iii*) heating an ultrathin Tc film to room temperature induces a transition to a meta-stable phase with lower density and intermediate molecular inclination angles. Finally, we discuss our data with regard to controlling the molecular inclination at Si-Tc interfaces.

## 2. Materials and methods

### 2.1. Sample Preparation

The substrates, RCA-cleaned c-Si(111) with a native oxide or PECVD-grown a-Si:H layers on glass substrates, were cleaned by short immersion into dilute hydrofluoric acid (1%, 2 min) and subsequently transported and loaded under N$_2$ into the UHV system. The air exposure during sample mounting was ≤ 1 min. Tc deposition was conducted in-system, with no break in UHV conditions between deposition and measurement. Tc films were deposited in the preparation chamber from a Knudsen cell at rates of 0.13–0.5 nm/min as calibrated with a quartz crystal microbalance. The substrates were cooled during deposition and also during measurement in the analysis chamber. A brief increase in cooling power was applied prior to transfer between the two chambers to ensure that no positive spike in temperature occurred.

The Tc deposition conditions are summarized in Table 2. Typically, Tc deposition was carried out at T$_{sub}$ = 265 K. One sample, H-Si(111) No. 1, was annealed to T = 720 K for 70 min in the UHV chamber to minimize the presence of adventitious carbon contamination [cf. Fig. 7 a)], then cooled to T = 200 ± 50 K[1] for Tc deposition. The Tc film for these conditions was ordered but showed no sign of interlayer diffusion. Its growth mode thus differs qualitatively from that of samples held at T = 265 K during growth. We denote this deposition regime as low temperature (LT).

### 2.2. NEXAFS and XPS

The experiments were performed at the HE-SGM beamline at HZB's synchrotron radiation facility BESSY II in Berlin. The polarization factor of the p-polarized X-ray beam, P, was 0.91. NEXAFS spectra were measured in partial electron yield (PEY) mode with a retarding voltage of −150 V using a custom-made multichannel detector. The X-ray incidence angle ($\theta$) is defined as the angle between the wave vector of incident X-rays and the surface plane of the substrate [see inset in Fig. 8 a)]. The X-ray photon energy scale in the NEXAFS experiments was calibrated by means of a carbon contaminated gold mesh upstream from the sample position that exhibit a characteristic and previously calibrated absorption peak [29, 30]. The raw PEY signal was corrected for the known spectral intensity distribution of the HE-SGM beamline. X-ray photoelectron spectroscopy (XPS) data were collected with a Scienta R3000 hemispherical electron analyzer employing 700 eV and 153 eV photon energies for survey and valence electron regions, respectively.

### 2.3. Theoretical Methods

For an as realistic as possible modeling of the investigated Tc monolayer structures on H-Si(111) surface, the molecular species together with the substrates have been calculated using periodic boundary conditions. The surface has been modeled by a slab containing four Si bilayers separated by a vacuum with a thickness of 25 Å. The bottom atoms were terminated by H atoms and frozen at the ideal lattice sites. The adsorbate structures have been relaxed using density functional theory (DFT) calculations with the Quantum ESPRESSO package (QE) [31]. The PBE functional [32] complemented with dispersion correction (DFT-D) [33] was used to model the electron exchange and correlation. Norm-conserving pseudopotentials were used to describe the electron-ion interactions. As basis functions, plane waves up to a cutoff energy of 60 Ry were used. Convergence criteria of $10^{-4}$ eV/Å for the maximum final forces and $10^{-8}$ eV for the total energy were used. The gauge including (GI-PAW) pseudopotential scheme was used to calculate the spectroscopic X-ray fingerprints of the Tc I and Tc II crystal structures as well as different adsorbate structures on H-Si(111). Specifically, we employed the Xspectra code [34, 35] of QE to calculate the NEXAFS C K-edge. To model the 1s core hole, scalar-relativistic multiprojector GIPAW pseudopotentials with a corresponding occupation of the inner shells were generated. This approach yields reliable results [36, 37], and is computationally very efficient in describing excitonic effects [38].

The 39 recursion scheme was employed to expand the Green's function of the empty states (continued-fraction expansion) [39-41], allowing for the calculation of the XAS cross sections. By employing this procedure, the near- and far-edge features of the absorption spectrum can be reliably described [42-45]. As quadrupole contributions are very small and restricted to the pre-edge region, we focused on the dipole-contributions. The XAS cross-sections of each carbon atom were calculated in a 40 eV range above the respective K-edge. Then, the total NEXAFS spectrum of the molecular system was obtained from the superposition of the individual spectra offset by the corresponding core-level-shifts (CLS), calculated within DFT [46-49] for each carbon atom (see e.g., Refs. [36, 37, 50] for further details). For structures with axial symmetry, the XAS cross section ($\sigma_{XAS}$) for an *E*-field component, tilted by an angle $\theta$ with respect to the surface normal, can be accurately calculated by averaging over two orthogonal azimuthal orientations (for an arbitrary angle $\varphi_0$):

$$\sigma(\theta) = 1/2 \ [\sigma_{XAS}(\theta, \varphi_0) + \sigma_{XAS}(\theta, \varphi_0 + 90°)]$$

---

[1] The thermocouple was positioned without direct contact to the sample, so that the actual $T_{sub}$ remains less defined than for the other samples. $T_{sub}$ should be substantially lower than the 265 K registered by the thermocouple, but still above the transition between disordered and upright-standing molecules, which we expect somewhat below $T_{sub}$ = 200 K observed for AlO$_x$/Ni$_3$Al(111) [9].

In the present case, the threefold symmetry of the substrate is broken by the attached molecules. To account for the enhanced variation with $\varphi$, we thus considered at least four different azimuthal cases (also $\varphi_0 + 45°$ and $\varphi_0 + 135°$) while determining the angular averaged spectra. The spectra were then aligned to the respective experimental K-edges. The linewidth was energy-dependent, [51] rising arctan-like from 0.1 eV at 280 eV to 1.0 eV at 320 eV.

## 3. Results

### 3.1. DFT-analysis of molecular structure of first monolayer

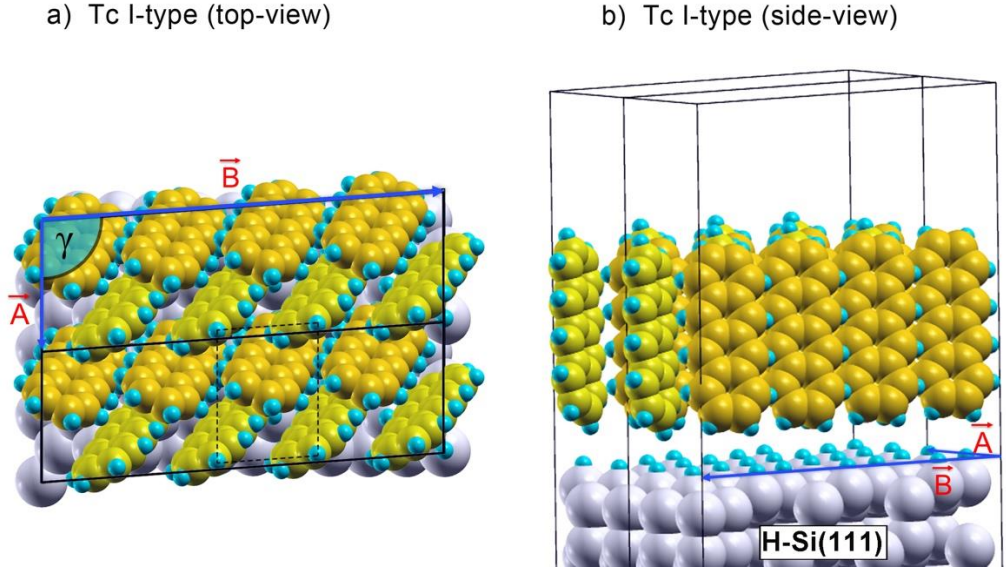

**Figure 3:** Modelling an ideal Tc monolayer (in LT regime): a) and b): top- and side-views, respectively, of the DFT-calculated *most stable* monolayers of Tc molecules at H-Si(111), denoted as Tc I-type monolayers. $\vec{A}$, $\vec{B}$, and the angle $\gamma$ define the lateral supercell used to model the structure: $\vec{A} = 0.772$ nm, $\vec{B} = 2.370$ nm and $\gamma = 94.21°$ resembling 1×4 repetition of the crystalline unit cells (dashed lines). The corresponding epitaxy matrix is given by $\begin{pmatrix} A_{TcI\_type} \\ B_{TcI\_type} \end{pmatrix} = \begin{pmatrix} 2 & -2 \\ 3 & 4 \end{pmatrix} \begin{pmatrix} x_1 \\ x_2 \end{pmatrix}$, where $\vec{x}_1$ and $\vec{x}_2$ indicate the primitive vectors of the Si(111) surface. The carbon atoms of the two non-equivalent Tc molecules within a unit cell are shown in slightly different colors.

We start our theoretical investigations by employing DFT total-enegy calculations to determine the most stable geometries of an ideal Tc monolayer on H-Si(111). Since the experiments do not provide an atomic registry between the Tc film and the substrate lattice, a variety of possible surface supercells with different size, shape, and molecular coverage within each cell, have been probed. Within each of the calculated structures, we allowed the starting configurations of Tc to have different possible inclination angles with respect to the surface normal and different azimuthal orientations with respect to its high-symmetry directions before they were completely relaxed.

According to our total energy calculations, the most stable monolayer structure features properties close to the Tc I (bulk) phase, whereby eight Tc molecules are included within a (1×4) surface cell (see Fig. 3). In the following, we denote this structure as Tc I-type. Furthermore, to simplify the discussion of the lateral dimensions, we denote the area of the $\vec{ab}$-plane of the Tc I unit cell [see Fig. 2 b)] as $A_{TcI}$ and use it as reference for all other structures. Apart from the 1×4 reconstruction, the density of the Tc I-type surface cell (dashed lines in Fig. 3) is close to that of the crystalline Tc I unit cell and the area per molecule is 97% of $A_{TcI}$. The two non-equivalent types of Tc molecules show the characteristic aspects of the Tc I phase (one molecule is 'simply' tilted, while the second, shown in slightly lighter colors, is tilted 'over the edge') and the angles $\alpha_{LA} = 67.5°$ and $\alpha = 73.5°$ are close to, but still distinct from those derived from the XRD data for crystalline Tc I (bulk) [24] ($\alpha_{LA} = 69.5°$ and $\alpha = 74.5°$). Such differences have to be expected, since the modelled monolayer experiences a weak, but direct interaction with the substrate, not present in the Tc I (bulk) unit cells. The total-energy calculations yield an adsorption energy of -1.27 eV (per Tc molecule) for the Tc I-type adsorbate structure, which is higher by 0.52 eV than the adsorption energy for a single flat-lying molecule (see also Table 1). In comparison with a single upright standing Tc even 1.07 eV are gained showing that intermolecular interaction is predominantly responsible for the stability of the Tc I-type monolayer.

**Table 1:** Adsorption energies per a Tc molecule ($E_{ads}$) calculated for the structures shown in Figs. 4, 3, 6. *Note that the so-called on-side geometry is only stable assuming C1h symmetry with the molecular plane as mirror plane, preventing the structure from 'falling down' to a flat-lying geometry. All ML consist of rather upright standing molecules. A comparison with the respective single Tc absorption shows that the stability of the ML is predominantly (70…85%) due to intermolecular interaction.

| Structure | | $E_{ads}$ [eV] |
|---|---|---|
| single Tc molecules on H-Si(111) | flat-lying | -0.75 |
| | on-side | -0.35* |
| | upright standing | -0.20 |
| ML of Tc molecules on H-Si(111) | Tc I-type | -1.27 |
| | Tc II-type | -1.20 |
| | Tc Ib-type | -1.10 |

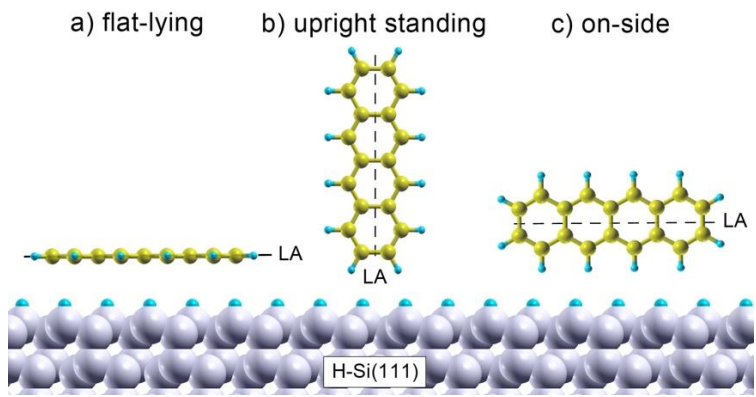

**Figure 4:** Variety of possible adsorption geometries of single Tc on H-Si(111). In the a) flat-lying [b) upright standing] geometry, both the long molecular axis (LA) as well as the molecular plane are parallel [perpendicular] to the surface plane. In the on-side geometries, the molecular plane is perpendicular to the surface plane, while LA is parallel to it. For further details, see also Table 1.

In the following, we study the structure of an ideal Tc monolayer as a function of the lateral dimensions of the unit cell. In other words, we report on the changes of the Tc inclination upon variation of the local molecular coverage. For this purpose, we assess the related changes of the total energy and molecular orientation within a Tc monolayer. Since we have to compare our results with the NEXAFS measurements, besides $\alpha_{LA}$ we report also the value of the NEXAFS inclination angle $\alpha$. In order to circumvent misleading, non-continuous strain effects within the substrate, these calculations are performed initially without substrate and belong, thus, to a completely unbound (free) Tc-monolayer. In the light of our total energy calculations (see Table 1), this appears to be a reasonable approximation, at least for dense monolayers. This is further supported by the fact that the minimum total energy structures with and without substrate are very similar. For the free monolayer, the total energy minimum is still found around 97 % of $A_{TcI}$, while providing $\alpha_{LA} = 68.5°$ and $\alpha = 73.8°$, in nearly perfect agreement with the values (67.5° and 73.0°) calculated for the Tc I-type monolayer at H-Si(111), see also Fig. 3. The Tc-substrate interaction only weakly affects the molecular inclination, and does not change the calculated trend considerably.

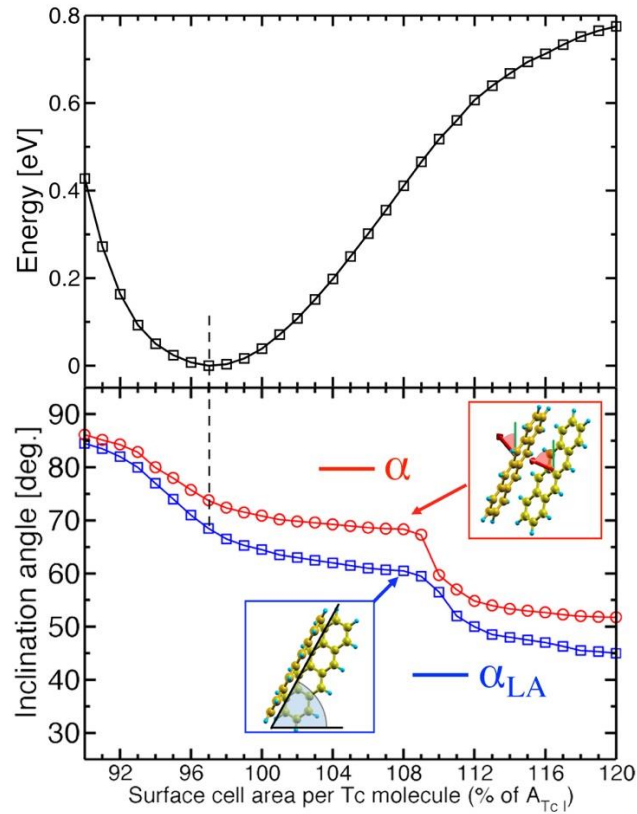

**Figure 5:** (*top*): Total energy for Tc monolayers as function of the surface unit cell area per molecule, referenced to that of bulk-Tc ($A_{TcI}$). The energy minimum is at 97% of $A_{TcI}$, as in case of the Tc I-type structure shown in Fig. 3. (*bottom*): Corresponding geometric inclination angle $\alpha_{LA}$, defined by the average orientation of the molecules' long axis (LA) against the surface plane. Note that $\alpha_{LA}$ is different from $\alpha$ determined from the NEXAFS analysis, which is also included. For increasing lateral cell size to about 130%, $\alpha_{LA}$ goes slightly down to about 40°, before for even lower coverage more individual arrangements in pairs or even flat-lying molecules are preferred.

Fig. 5 shows the calculated inclination angles as a function of unit cell size. Both angles, $\alpha_{LA}$ as well as $\alpha$, decrease with increasing cell size, i.e. as expected, the Tc molecules tend to be more planar for locally lower molecular coverages.[2] In the range between 90 % and 108 % $A_{TcI}$, the decrease of $\alpha_{LA}$ and $\alpha$ is almost *arccot*-like, smoothly approaching $\alpha_{LA} = 60°$ for 108%. Notably, this flattening of the molecules reflects the effect experimentally observed [25, 11]. We will come back to this point later in the experimental part. Around 110 % surface area, a qualitative change of the molecular arrangement yields a sudden decrease of the inclination angle by about 10° allowing the Tc molecules to adapt more planar geometries. Decreasing the lateral dimensions to about 90% of $A_{TcI}$, on the other hand, yields more inclined molecules which resembles those of Tc II (thin film) structures.

To explore the influence of the substrate, we determined the stable structures of Tc molecules in supercells of lateral dimension larger and smaller than the Tc I-type monolayer by about 10%, whereby the lattice vectors have to be chosen to fit to the underlying substrate. In both cases the molecules are again arranged in a herringbone structure. Figure 6 a, c) shows a surface cell forming a 1×5 repetition of unit cells ($\vec{A} = 0.772$ nm, $\vec{B} = 2.702$ nm, $\gamma = 90°$), where the area per molecule is now 89% of $A_{TcI}$. It includes ten upright standing Tc molecules ($\alpha_{LA} = 88°$ and $\alpha = 88.2°$) and resembles the thin film phase [20]. This monolayer structure is thus denoted as Tc II-type. Figure 6 b, d) provides an example for a less dense arrangement, whereby the area per molecule is 110% of $A_{TcI}$. It shows a simple, rectangular 1×1 surface cell of dimensions $\vec{a} = 0.675$ nm, $\vec{b} = 0.772$ nm, and includes two Tc molecules. Similar to the Tc I-type it resembles the Tc-bulk molecular arrangement, but without adapting an oblique unit cell ($\gamma = 90°$). It will be thus denoted as Tc Ib-type in the following. In comparison with the Tc I-type, the Tc molecules in the less dense Tc Ib-type structure show a smaller inclination angle $\alpha_{LA} = 63°$. Since the angle between the conjugated molecular plane and the surface normal of the second Tc molecule (which is tilted 'over the edge') is about 76°, the average NEXAFS $\alpha$ for this structure is about 68.5°, again considerably smaller than the value obtained for the Tc I-type arrangement.

---

[2] Previous studies reported that a decrease of the local coverage enhances the importance of the molecule-surface interaction and favors smaller molecular inclinations [56-58].

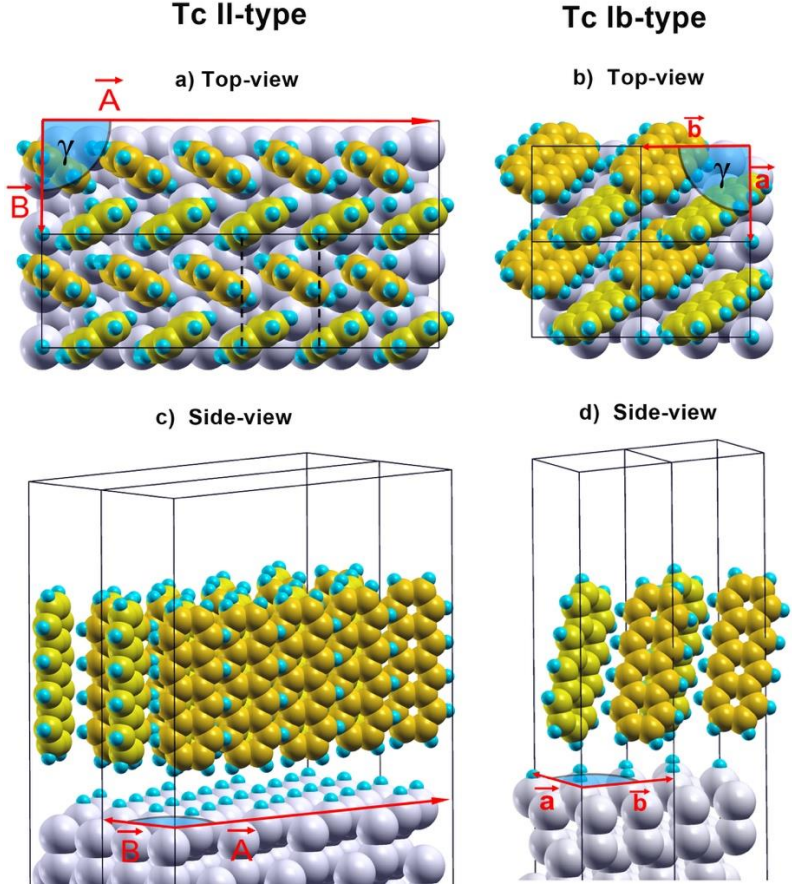

**Figure 6:** *a*) [b)] Top and c) [d)] side-views of the geometries of a Tc monolayer calculated in rectangular supercells of lateral dimensions $\vec{A}$ = 0.772 nm, $\vec{B}$ = 2.702 nm [$\vec{a}$ = 0.675 nm, $\vec{b}$ = 0.772 nm] and $\gamma$ = 90°, resembling 1×5 [1×1] unit cell reconstructions for Tc II-type [Tc Ib-type], respectively. These structures represent 10% more dense (Tc II-type) and 10% less dense coverage (Tc Ib-type) in comparison to the most stable Tc I-type structure shown in Fig. 3. The corresponding epitaxy matrices are given by $\begin{pmatrix} A_{Tc\,II\_type} \\ B_{Tc\,II\_type} \end{pmatrix} = \begin{pmatrix} 2 & -2 \\ 4 & 4 \end{pmatrix} \begin{pmatrix} x_1 \\ x_2 \end{pmatrix}$ and $\begin{pmatrix} a_{Tc\,Ib\_type} \\ b_{Tc\,Ib\_type} \end{pmatrix} = \begin{pmatrix} 1 & 1 \\ 2 & -2 \end{pmatrix} \begin{pmatrix} x_1 \\ x_2 \end{pmatrix}$, where $\vec{x}_1$ and $\vec{x}_2$ again indicate the primitive vectors of the H-Si(111) substrate.

The total-energy calculations yield an adsorption energy per Tc molecule of -1.20 (-1.10) eV for Tc II (Tc Ib)-type structures, as expected slightly lower than for the Tc I-type monolayer (-1.27 eV). This indicates that the formation of the Tc I-type monolayer on H-Si(111) is most favorable. However, in case of the Tc-II type structure, this assessment is reverted when including its higher areal density. The adsorption energy per nm$^2$ is 0.578 meV and, by this, ~4% larger than for the Tc-I type monolayer (0.555 meV). This effect suggests that kinetic effects during and after film growth [26, 59, 62] will influence whether preferentially Tc I or Tc II-type (thin film-like) molecular arrangement form. In addition, at elevated temperatures, vibrational and configurational entropy contributions to the adlayer free energy will modify the energetics. Their explicit treatment is beyond the scope of this work, but temperature effects can be implicitly taken into account via their influence on molecular coverage. In the following sections, we investigate the crucial role of the temperature to initiate structures deviating from the most stable ground state configuration, so that arrangements with intermediate molecular inclinations like, e.g., Tc Ib-type become possible.

## 3.2. Films grown and measured at LT

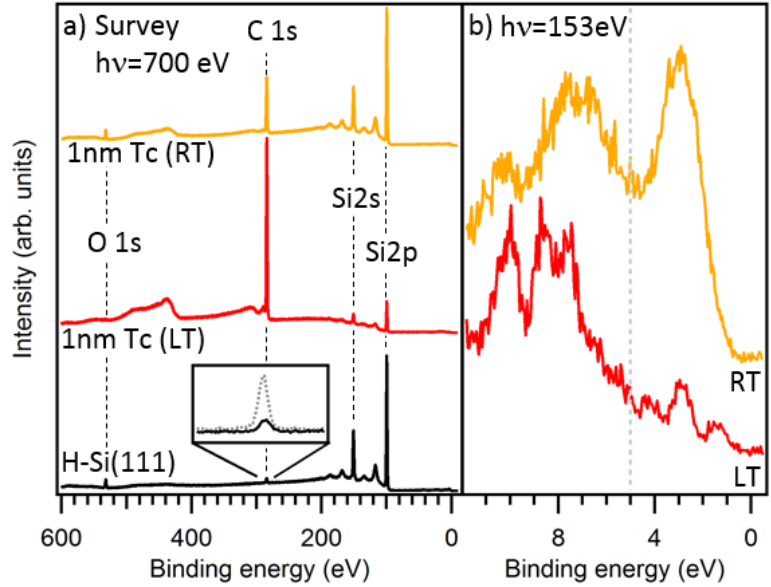

**Fig. 7**: a): XPS survey spectra of H-Si(111) before and after Tc deposition at LT as well as upon warming the Tc to room temperature (RT). The inset shows a close-up of the C 1s region and includes a spectrum of the substrate before annealing (gray dotted line). The annealed H-Si(111) substrate (720 K for 70 min, black) still features C 1s and O 1s intensity from substrate contamination not completely removed by the annealing. The red spectrum belongs to a Tc film of nominally 1 nm coverage deposited and measured at LT; the yellow spectrum was measured after warming up to room temperature (RT). Note (*i*) the increase of the C 1s signal upon Tc deposition and the corresponding decrease in Si 2p and Si 2s, and (*ii*) the opposite trend upon raising the temperature to RT. b): The valence electron region for the same Tc film as in a). For the cooled sample (red), the spectrum is dominated by the Tc molecular orbitals. Upon exposure to room temperature (yellow) the spectrum becomes dominated by the Si bands. The spectral regions above and below 5 eV (indicated by the vertical dotted line) were acquired with different statistics.

Table 2: Summary of the preparation and measurement conditions of the investigated samples and the film properties derived from XPS and NEXAFS. The minimum film thickness ($d_{min}$) is corrected by subtracting the $d_0$-value of the residual substrate-related carbon species. Both values were derived according to established procedures [52-54]. A systematic error of 30% is estimated due to the large uncertainty of the electron attenuation lengths in COMs [55] and the fact that our analysis relies on the ratio of two core levels with a 135 eV difference in binding energy. The fitted inclination angle of Tc ($\alpha$) based on NEXAFS analysis is also given. Note that due to the lower $T_{sub}$, H-Si(111) (No. 1) exhibits a different growth mode than the other samples. In addition, that the H-Si(111) (No. 2) with nominal 3 nm coverage likely corresponds to a different film thickness regime than the other films (c.f. Fig. 1)

| Sample | Nominal coverage (nm) | Rate (nm/min) | $T_{sub}$ (K) | $d_0$ (nm)/ $d_{min}$ (nm) | $\alpha$ (°) |
|---|---|---|---|---|---|
| H-Si(111) (No. 1) | 0 | - | RT | 0.05 / 0.00 | |
| H-Si(111) (No. 1) | 1 | 0.38 | LT | 3.30 / 3.25 | 76.3 |
| H-Si(111) (No. 1) | 1 | - | RT | 0.55 / 0.50 | 62.7 |
| H-Si(111) (No. 2) | 0 | - | RT | 0.20 / 0.00 | |
| H-Si(111) (No. 2) | 1 | 0.20 | 265 | 1.35 / 1.15 | 77.8 |
| H-Si(111) (No. 2) | 2 | 0.45 | 265 | 0.70 / 0.50 | 78.5 |
| H-Si(111) (No. 2) | 3 | 0.50 | 265 | 0.80 / 0.60 | 76.6 |
| a-Si:H | 0 | - | RT | 0.30 / 0.00 | |
| a-Si:H | 1 | 0.13 | 265 | 1.60 / 1.30 | 76.3 |
| a-Si:H | 2 | 0.50 | 265 | 1.35 / 1.05 | 77.5 |

With the theoretically predicted characteristic angles as a function of local Tc coverage at hand, we set out to determine experimentally the orientation of the first monolayer as a function of coverage and temperature. We first look at Tc/H-Si(111) grown with $T_{sub}$ sufficiently high for an ordered film to be formed and sufficiently low to prevent any significant dewetting, thus resulting in a (in the best case) closed and ultrathin Tc film. In a first preparatory step, we determine the background signal of the H-Si(111) substrate after annealing at 450 °C for 70 min in the vacuum chamber. The black spectrum in Fig. 7 a) presents a XPS survey spectrum of the pristine H-Si(111) substrate immediately after the annealing procedure. Besides the expected Si lines, also C signals are visible. The inset in a) shows the C 1s region and includes the measurement before annealing (dashed gray line). Upon annealing, the carbon contamination was reduced by about 80%. This sample was also used for the measurements with the film at room temperature presented later. All other substrates did not undergo any in-vacuum annealing and, thus, contain the full substrate-related contribution.

The annealed substrate was then cooled to LT for Tc film growth and subsequent analysis. A Tc film of 1 nm nominal coverage was deposited onto H-Si(111) and then transferred to the analysis chamber without significantly raising the sample temperature. As shown in Fig. 7 a), the Si 2p and 2s signals were greatly reduced following Tc deposition, while the C 1s signal was increased. In the valence band region, shown in Fig. 7 b), features derived from Tc's molecular orbitals dominate the spectrum. The effective attenuation length (EAL) in COMs for electrons with 153 eV kinetic energy is 0.6–1.0 nm [52, 55]. Accordingly, the observed suppression of Si-related spectral intensity to < 10% its original value implies that a closed Tc film of at least 1.4 nm height was formed. Since this value is slightly larger than the nominal coverage, we conclude that the substrate temperature LT effectively suppressed Tc dewetting.

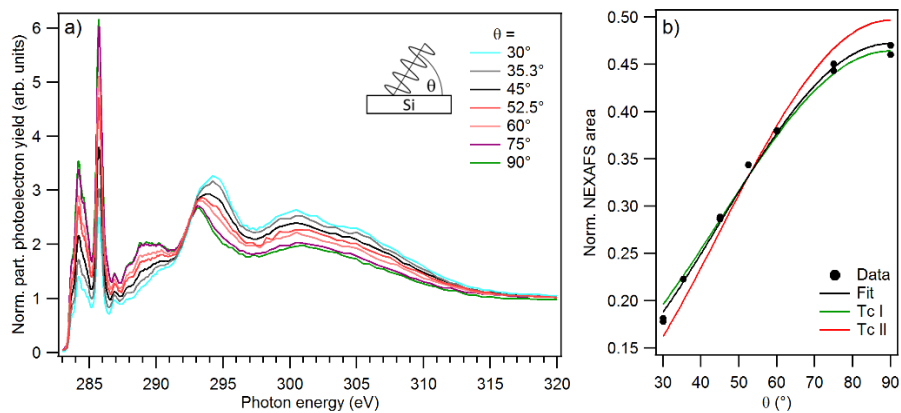

**Figure 8**: a): C K-edge NEXAFS spectra (for different angle of incidence $\theta$, see inset) for a nominally 1nm-thick ultrathin Tc film at H-Si(111) grown and measured at LT. b): The $\pi^*$ resonance intensities derived from the spectra shown in a) as a function of $\theta$, a fit to the data according to Eq. (1), and the expected curves for pure Tc I (bulk) and Tc II (thin film) crystal phases. (Error bars are within the size of the symbols).

To characterize the spectral signature and the conformational properties of the LT Tc film, C K-edge angular resolved NEXAFS measurements were performed. The normalized spectra taken at different incidence angles $\theta$ are presented in Fig. 8. Their overall shape is comparable to previous reports [10, 11, 25, 60]. The apparent strong dichroism in the spectra shows that the film exhibits a preferential molecular alignment. The spectral contribution in the region below 290 eV stems exclusively from $\pi^*$-transitions. Above 293 eV, the spectrum is dominated by $\sigma^*$-transitions. The opposite behavior, i.e., increase (decrease) of the peaks intensities within the $\pi^*$ region ($\sigma^*$ region) for larger incident angles suggests molecules closer to an edge-on (either side-on or upright-standing) adsorption geometry.

For further analysis, we determined the incidence-dependent $\pi^*$-related main peak intensity $I(\theta)$ by fitting the spectral region below 286.5 eV with the following model [61][3] that assumes at least three-fold axial symmetry:

$$I_\alpha(\theta) = 1/2\,(3P\cos^2\theta - 1)\cos^2\alpha + 1/2(1 - P\cos^2\theta) \quad (1)$$

As a result, for a given inclination angle $\alpha$, $I_\alpha(\theta)$ becomes a function of $\theta$ alone. Fig. 8 b) shows the experimental data and the corresponding fit according to Eq. 1, yielding $\alpha = 76\pm1°$. To visualize the significance of this result, Fig. 8 b) also includes the expected curves for Tc I (bulk) [24] and Tc II (thin film) [20]. Taking into account the experimental error margins, the nominally 1 nm-thick *ultrathin* Tc film at H-Si(111) grown and measured at LT is in very good agreement with the Tc I (bulk)-derived curve, much better than with the curve for Tc II (thin film), rendering the inclination of the first Tc monolayer to resemble that of Tc I (bulk). This observation is in line with the DFT total energy calculations (reported in the previous section) that yield a Tc I-type monolayer ground state structure (see Fig. 3).

### 3.3. Films grown and measured at $T_{sub}$ = 265 K

We prepared Tc thickness series on H-Si(111) as well as on a-Si:H at $T_{sub}$ = 265 K. This temperature resulted in partial dewetting during growth of the films, but not during characterization. The thickness series consisted of a sequence of Tc depositions and X-ray measurements performed contiguously on the same sample, with nominal coverages of 1, 2 and 3 nm on H-Si(111) (1 and 2 nm on a-Si:H). Compared to the LT film, clear changes in the Tc-related XPS and NEXAFS signals were observed (see Fig. 9), but no clear influence of the coverage on $\alpha$ can be derived (see Table 2). To derive the growth mode, we look at the coverage-dependent substrate and Tc signals measured by XPS and NEXAFS. Throughout the whole series, the relative intensity of the substrate-related signals (both in the valence electron and C K-edge) are much more intense compared to those related to the lower substrate temperature (cf. section 3.2) suggesting different growth modes.

Concerning the film of 1 nm nominal coverage and assuming a homogenous thickness of Tc films, a lower limit $d_{min}$ for the Tc film thickness can be estimated from the Si 2s to C 1s XPS intensity ratio, resulting in 1.15±0.4 nm and 1.3±0.4 nm for H-Si(111) and a-Si:H, respectively (see Table 2). Due to Tc's complex growth mode [25, 26, 62], these $d_{min}$ values have to be analyzed with great care with respect to their evolution with the nominal coverage. Fig. 9 presents the valence electron and NEXAFS C K-edge spectra, showing Tc- and substrate-related spectral weights of comparable magnitude. As expected, the Tc features become more prominent upon increasing the nominal Tc coverage. But the derived $d_{min}$-value, in contrast, decreases. This inverted behavior holds for both investigated substrates and can be explained by local dewetting, well-known for COMs with strong dewetting tendency [16]. If we neglect desorption, the decrease of the overlayer signal upon increasing coverage means that areas with Tc thickness > 1.7 nm grow at the expense of areas of lesser thickness. Simultaneously, the proportion of surface area with film thickness ~1 nm also increases.[4] This points to a growth mode where significant dewetting of molecules from multilayer islands is induced, while, at the same time, the wetting layer of the film is much less affected by the dewetting and effectively grows in area.

---

[3] Note that Eq. 1 is not linear in $\alpha$ [30] and therefore the combined inclination angle $\alpha$ is not simply the arithmetic average of the individual values (69.48° and 81.87°) for the two molecules in the Tc bulk unit cell. The average would be 75.675° (instead of 74.5° resulting from Eq. 1).

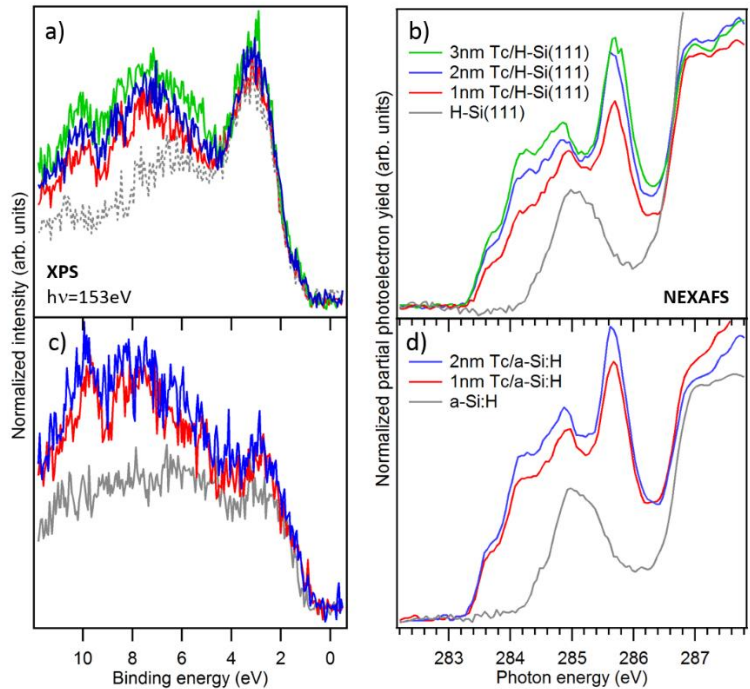

**Figure 9**: Experimental spectra for the Tc thickness series on H-Si(111), deposited and measured at $T_{sub}$=265 K. a): Valence electron region; b): C K-edge NEXAFS. $\theta$ was 20° (30°) for the Tc films (the substrate-only spectrum). c,d): The same as a,b), but for a-Si:H substrate. All spectra have been scaled so that the intensity of the substrate contribution stays identical. The spectral contributions of the substrate and the Tc film were determined via fits and difference spectra (not shown). The substrate-only spectrum in a) was not measured but simulated by subtracting a Tc-dominated spectrum (grown at LT and shown in Fig. 7) from the 1 nm Tc film.

3.4. Measurements at room temperature (RT)

Last but not least, we look at an ultrathin Tc film at $T_{sub}$ = RT in order to resolve possible temperature effects on the Tc orientation. To assess the corresponding film structure, we let the as-deposited film (grown at LT) thermalize to RT over the course of 9 hours before re-analyzing it.

---

[4] The given values follow from an analysis of the mean escape depths (MED) [52] for XPS/NEXAFS with different energies, entering the analysis of the experimental intensity ratio: Photoelectrons from the Si 2s core level excited with $h\nu$ = 700 eV have a kinetic energy of 550 eV and a $MED_{550eV}$ of 1.7 nm. In contrast, photoelectrons from the valence electron region excited with $h\nu$ = 153 eV have a $MED_{153eV}$ of only 0.6 nm. C K-edge NEXAFS is sensitive to electrons with maximum kinetic energy of 285 eV resulting in a $MED_{285eV}$ of 1 nm. In the employed partial electron yield mode, a fraction of the inelastically scattered electrons is also detected, enhancing the MED, but only to limited extend.

Compared to the as-deposited film, the Si 2p (C 1s) signal in the XPS survey spectrum [see Fig. 7 a)] of the annealed film is significantly increased (decreased). The valence electron region is now clearly dominated by Si features [see Fig. 7 b)]]. These observations clearly show that at least parts of the RT-exposed H-Si(111) surface are now dramatically less covered by Tc molecules.

Further support for this scenario is given by the corresponding NEXAFS spectra [see Figs. 10 a) and b)]. A proper description requires much larger background contribution (BG) than those for the 265 K films. Since the latter films do not show significant spectral changes during several hours, the qualitative different Tc spectral weight cannot be explained by an aging effect. Instead we have to deduce that dewetting and potentially also desorption is significantly more pronounced for the room temperature film. This sample indeed clearly shows a continued decrease of the Tc spectral weight (40% in 100 minutes), evidencing Tc dewetting/desorption almost linearly in time. This trend is accounted for in the data presented in Fig. 10 c) that yields $\alpha = 47.5\pm3°$ for the background signal (BG, originating from the carbon contamination of the substrate) and, most importantly, $\alpha = 63\pm2°$ for the Tc film. This is in line with previous reports in literature [10, 25] and also in accordance with the reduction of the inclination angle in less dense covered regions predicted by the DFT-calculations shown in Fig. 5. We will come back to this fact in section 4.

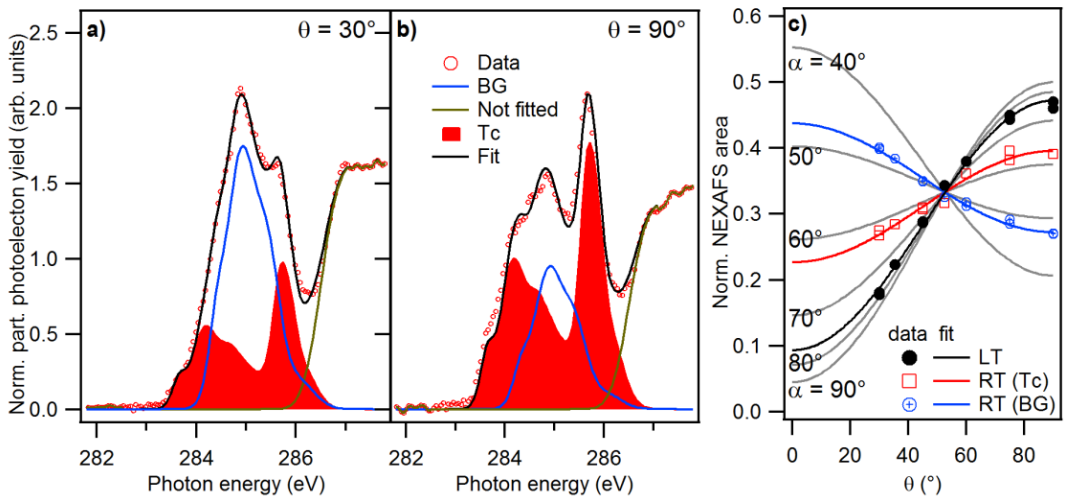

**Figure 10**: a, b): NEXAFS C K-edge spectra measured at $\theta = 30°$ and $\theta = 90°$ X-ray incidence, respectively, for 1 nm Tc at H-Si(111) grown at LT and heated to RT. The fitted spectral contributions from Tc and the substrate are also shown. c): $\pi^*$ resonance intensities as a function of $\theta$ for the following signals: Tc measured at LT, Tc measured at RT, the residual substrate carbon (background, BG) signal at RT. Data shown with fits according to Eq.1, together with curves expected for some indicated $\alpha$.

3.5. Calculated NEXAFS C K-edge for Tc adsorbate models

For further structure assessment, we provide DFT-calculated angle-dependent NEXAFS C K-edge spectra for reasonable model structures and compare them with the experimental spectra for the ultrathin film grown at LT. NEXAFS C K-edge spectra of single Tc molecules in '*gas phase*' (i.e. without intermolecular interaction) have been calculated before by Sueyoshi *et al.* [60] and Fratesi *et al.* [63]. The present calculations go beyond single-molecule spectra. Our NEXAFS spectra calculated for extended, 2D periodic structures, i.e. for the Tc I-type adsorbate and Tc I/Tc II crystalline structures, account for the influence of the substrate and intermolecular interactions. The latter turn out to be predominantly responsible for the stability of the molecular Tc monolayer (see also section 3.1 and Table 1).

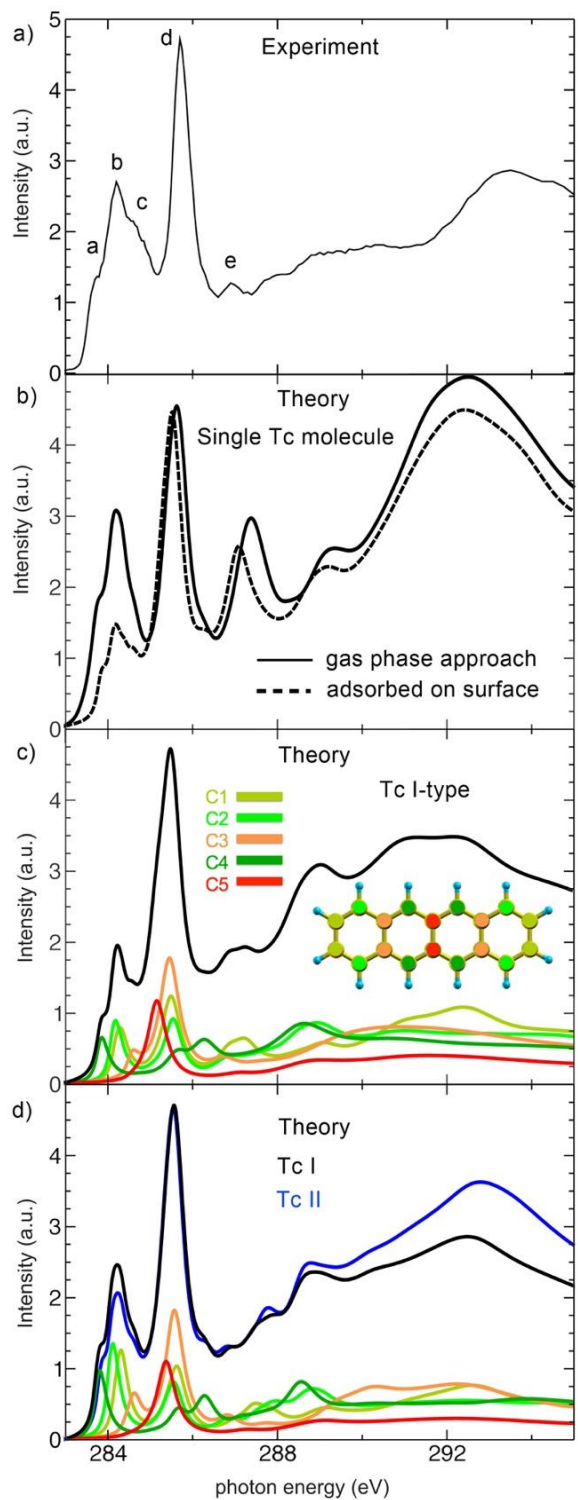

**Figure 11**: a): Experimental NEXAFS C K-edge spectrum of the LT film, shown exemplary at the incident angle $\theta$ = 52.5°. b), c) and d): DFT-calculated NEXAFS C K-edge spectrum for b) single Tc molecule in gas phase/vacuum (solid line) and adsorbed on surface (dashed), c) the most stable Tc monolayer structure (Tc I-type shown in Fig. 3) and d): the crystalline bulk-phase (black) and the crystalline thin-film phase (blue), respectively. All calculated spectra are also shown at incident angle $\theta$ = 52.5°. In c) and d), the contributions of different C-atom types to the overall spectrum of the Tc I-type and Tc I (bulk) are also shown (bottom). The inset shows the chemical structure of Tc and serves as reference for the color code.

Fig. 11 a) shows the experimental C K-edge NEXAFS spectrum for the LT film taken under incident angle $\theta = 52.5°$ [c.f. Fig. 8 a)]. In the energy range up to 286 eV (the $\pi^*$ region), the first four peaks *a*, *b*, *c*, d are found at 283.75, 284.2, 284.55, and 285.7 eV, respectively. In the energy range above 286 eV, several broad resonances are superimposed to the ionization continuum forming a continuous structure of the spectra. In the energy range 287-292 eV, the spectrum shows almost constant intensity, while the intensity of the peak at 293.3 eV is about 60% of the main peak *d*.

Fig. 11 b) shows the corresponding calculated NEXAFS C K-edge spectra of the single Tc molecule in gas phase (solid line). It reproduces theoretical results of Fratesi et al. [63], and shows reasonable agreement with the experimental data, at least up to 286 eV. Above 287 eV, however, the calculated spectrum deviates considerably from the measurements. Here, strongly pronounced resonances appear at energies 287.4 and 292.5 eV. Whereas their energy can be somehow related to experiment, their intensity is drastically overestimated. Fig. 11 b) (dashed line) shows the spectrum of a single Tc molecule adsorbed on the surface (upright standing geometry). The resulting spectrum still features in many aspects the gas phase spectrum. The peak positions are slightly shifted, but only the position of the e peak fits better to the experimental spectrum. The intensities of the *a*, *b*, *c* peaks, however, provide even larger discrepancies with the measurements, e.g., the calculated intensity ratio *b*/*d* is 0.33, while it is 0.58 in the experiment. Single molecule approaches with and without substrate are, thus, quite far away from being representative of the measured ultrathin Tc films. Obviously, the deviations obtained are not related to the influence of the substrate, but to the absence of the intermolecular interactions as it will be shown in the following.

Fig. 11 c) shows the spectrum calculated for the Tc I-type structure (see also Fig. 3), which accounts for both the substrate and intermolecular interactions. Indeed, the spectrum now shows better agreement with the experiment. The calculated energies of the characteristic *a-e* peaks coincide within 0.05 eV with the experimental data, and the intensity of the formerly overestimated *e* peak and the broad resonance around 292 eV are now correctly reproduced. However, the binding energies above 288 eV as well as the intensities of the *a*, *b*, *c* peaks are still underestimated; i.e., the calculated intensity ratio *b*/*d* is only slightly increased to 0.4. Whereas the deviations above 288 eV might be related to the employed semi-local exchange-correlation functionals [63] or to possible contributions from vibronic excitations to the final states, the deviations below 288 eV indicate some discrepancies between the modelled and measured structures. Our theoretical modelling assumes a perfect *single* monolayer of molecules in a direct contact with the substrate. The measured structure seems to deviate at least partially from this 1 ML assumption.

This argument is further supported by calculating the angle-dependent NEXAFS C K-edge spectra for the published XRD-derived crystal structures of bulk (Tc I) [24]. For comparison, the corresponding spectrum of thin-film Tc (Tc II) [20] is also shown. The calculated spectrum, especially for Tc I, is now in very good agreement with the experiment. The formerly underestimated characteristic peaks *a*, *b*, *c* are now almost identical to the measurements not only with respect to their binding energies, but also in terms of their shapes and intensities. Thereby, the first peak *a* results mainly from excitations of C4 carbon atoms into the lowest $\pi^*$ orbitals (see decomposition of the spectrum into contributions from the five non-equivalent (C1 to C5) types of C atoms). Peak *b* is dominated by excitations from C1 and C2 carbon atoms. The feature *c* includes contributions from C1, C2 and C3, whereas feature *d* carries contributions from all carbon types. Apparently, Tc I fits much better to the measurement than Tc II. A *b*/*d* intensity ratio of 0.53 (Tc I) fairly agrees with the experimental data (0.58), whereas the calculated ratio for Tc II (0.46) appears much too small. Furthermore, the intensity of the Tc II resonances in the range above 289 eV are dramatically overestimated, whereas the Tc I curve perfectly reproduces the experimental intensity ratio of about 60 % of the main peak *d*. All-in-all, this supports *(i)* again the assumption of considerably tilted, preferential bulk phase-like molecular orientation in the first Tc monolayer(s) and *(ii)* that the closed ultrathin Tc film grown and measured at LT does not provide an entirely single monolayer, but contain residual regions with thickness > 1 ML.

## 4. Comparative Discussion

### 4.1. Molecular structure of ultrathin Tc films grown at LT and 265 K

From our NEXAFS data, the presence of a planar-oriented interlayer of molecules like in the case of pentacene on clean Si(001) [14] and $\alpha$-Al$_2$O$_3$(0001) [64] can be safely excluded. Instead, the molecules in the ultrathin Tc film feature an orientation with average inclination angles between 76° and 78°. However, from NEXAFS alone it is not clear if the molecules are oriented with the molecular long axis (LA) closer to be parallel to the underlying surface or perpendicular to it (see Fig. 4). From all reports in literature (utilizing XRD, LEED, and AFM) [9, 10, 19, 20, 22, 62, 65, 66] as well as our own total energy calculations, we can assume more upright-standing Tc interlayer orientation with a thickness of 1.0 – 1.3 nm in our experiments. This means that the data for the film grown at LT (see Table 2) translates to an average thickness of 2.5±1 ML, i.e. a closed first ML, and the second ML at least predominately closed.

In contrast, the Tc films grown at $T_{\text{sub}}$ = 265 K showed clear signs of interlayer diffusion of multilayer molecules during growth (once formed, the films were stable for hours when measured at $T_{\text{sub}}$ = 265 K). Upon depositing additional Tc molecules onto films of nominally 1 nm, parts of the film with thickness ≥ 2 ML lose molecules to thicker areas and thus contribute less spectral weight.[5] At the same time, the first ML continues to grow with nominal coverage over the investigated thickness range, demonstrating the stability of the first ML with regard to dewetting at that temperature. Both trends translate to an increasing sensitivity of the surface-sensitive NEXAFS measurements to the desired uncovered first ML with increasing nominal coverage.

Within ~2°, all investigated ultrathin (1 – 3 ML thick) Tc-films for $T_{\text{sub}}$ = 265 K yielded the same $\alpha$–value of approximately 77°, which is also true for the Tc/H-Si(111) film at LT. No clear coverage-dependent $\alpha$-trend can be derived for the thickness series. Equally, the kind of substrate, crystalline H-Si(111) or amorphous a-Si:H, seems to have no clear influence on the molecular orientation. The $\alpha$ values do not fully agree with pure bulk-like Tc I phase (74.5°). However, within the calculated error margin the Tc I angle fits much better than that expected for a thin film (Tc II) phase (85.4°). This is in line with the better agreement of calculated NEXAFS spectra for the Tc I (bulk) crystal structures, and indicates that for as-deposited films, the orientation of the first ML clearly differs from that of the thin film regime. In other words, the ultrathin film regime, with 1 to ~3 ML, should be recognized as being distinct from the thin film regime (5 ML to 20 ML) [21]. In both transition regimes into the less inclined bulk-like Tc I phase, i.e. around 20 ML and for 3 – 4 ML, a mixture of both phases seems conceivable. Indeed, XRD measurements of Tc/H-Si(111) did resolve both phases for larger thicknesses [23].

---

[5] This behavior partly matches a previous report for Tc/SiO$_2$ that found that "with increasing deposition of Tc […] islands with less than three molecular layers are all subject to […] mass transports from lower layers to higher layers." [26].

4.2. Molecular orientation of Tc films after heating to RT

In the following we turn our attention to the film deposited at LT and later heated to RT. In this case, $\alpha$ is reduced to 63±2°, i.e., significantly lower than prior to annealing. This trend is opposite to that observed upon thermally activated erection of kinetically frustrated flat-lying polyacenes [9, 15, 16]. The resulting Tc inclination reflects several other observations in literature: upon stepwise Tc deposition onto H-Si(100) held at RT, $\alpha = 65\pm3°$ was derived from NEXAFS in the low coverage regime (nominally 1.2 ML) [25]. For > 3 ML coverage, $\alpha$ quickly converges to 78.4°, i.e., values similar to those observed for Tc films measured at $T_{sub} \leq 265$ K in the present work. Schedel *et al.* investigated Tc on oxidized Si(111) and reported the same trend towards lower $\alpha$ with decreasing coverage, but did not give a number for the low coverage regime [11]. For the (non-luminescent) disordered Tc films on $AlO_x$/$Ni_3Al$(111) grown at 100 K, upon annealing to 280 K, $\alpha$ of a multilayer film remains unaffected, but for 1 ML $\alpha$ changes from 43±5° to 63±5°.[10] The latter value is again identical to what we observed for the RT exposed sample in this work, and what was reported for RT grown Tc on H-Si(100) [25], suggesting a substrate-independent effect. Notably, the measurement of the annealed film on $AlO_x$/$Ni_3Al$(111) was performed with the sample again cooled to 100 K, indicating that the Tc structures with small inclination angle that emerge upon annealing are at least meta-stable.

All these findings indicate that at low coverages and ~RT, Tc molecules assemble into 1 ML islands that feature much smaller inclination angles than those known from the Tc I/II crystalline phases. A similar angle is reflected by the DFT calculated limit for Tc Ib-type, indicating less inclined, but still herringbone-like ordered (Tc-Ib type) monolayers (cf. Fig. 6 b,d) in the lowered coverage regime (see also Fig. 5). An accidental coincidence of all these $\alpha$ values is unlikely, and we have to take the $\alpha$ value of 63° seriously into account. Based on our DFT-analysis, we tentatively attribute the RT phase with $\alpha = 63°$ to a structure with an intermediate inclination similar to Tc Ib-type shown in Fig. 5 (bottom).

## 5. Conclusions

We elucidated the structure of ultrathin tetracene (Tc) films on H-passivated Si(111) and hydrogenated amorphous Si (a-Si:H) by combining NEXAFS and XPS experiments with DFT calculations. Our total energy calculations show that the stability of the Tc monolayer is predominantly due to intermolecular interaction, confirming that the lowest Tc layer is only weakly interacting with the substrate. This is also reflected by the calculated NEXAFS spectra. Whereas a periodic modelling of the molecular Tc layer is crucial for describing the experimentally observed spectra, the influence of the substrate is of minor importance. As a result, the gained knowledge on Tc-film growth can be safely transferred to any weakly interacting substrate and reasonably discussed based on published data obtained for similar substrates like H-Si(100) [25], oxidized Si(111) [11], or even $AlO_x$ [10].

In detail, ultrathin Tc films ($\leq 3$ ML) grown and measured at $T_{sub} \leq 265$ K are dominated by almost upright-standing, but still below $\alpha = 79°$ inclined molecules. The lowest monolayer, i.e. the molecules with direct contact to the substrate show a molecular packing almost identical to Tc bulk crystalline material, but with slightly different inclination angle (Tc I-type structure). The transition into the thin-film phase (Tc II) requires 2 – 4 ML molecular material in order to decouple the deposited molecules completely from the weakly-interacting substrate. Finally, at film thicknesses of about 20 – 100 ML the transition into the bulk-like Tc I phase takes place.

In contrast, ultrathin Tc films deposited at LT then heated to RT undergo a structural transformation, and exhibit a smaller $\alpha$ of about 63°. Based on a comparative analysis that includes observations for Tc on different weakly interacting surfaces, we suggest that both a low local coverage and a sufficiently high temperature are necessary conditions for the intermediate-inclination structure to appear, and that this structure is at least meta-stable. The larger overlap with the substrate wavefunctions makes this arrangement attractive for an optimized interfacial electron transfer in optoelectronic devices and solar cells. Hence, any condition (for example periodic heating and cooling or varying the growth rate) promoting the growth of more and larger islands, in the best case a perfect monolayer, with 63° inclined molecules appears to be beneficial for technological applications. We furthermore suggest following up the preparation of the RT-annealed ultrathin films by growing subsequent layers at decreased $T_{sub}$ to stabilize the 63° Tc orientation at the interface against the structural transition around 3 ML that was observed for RT-grown samples.

# Author Information

**Corresponding Authors**

*E-mail: jens.niederhausen@helmholtz-berlin.de

**E-mail: aldahhak@mail.uni-paderborn.de

‡These authors contributed equally.

**Notes**

The authors declare no competing financial interest.

# Acknowledgements

Numerical calculations were performed using grants of computer time from the Paderborn Center for Parallel Computing (PC$^2$). The Deutsche Forschungsgemeinschaft (DFG) is acknowledged for financial support via the collaborative research center TRR 142 – project number 231447078. We are indebted to Bundesministerium für Bildung und Forschung (BMBF) for funding part of this research through the SISSY project under project number 03SF0403 and the team of the Energy in situ Laboratory Berlin (EMIL) for help with the experiments. RWM thanks the Helmholtz Association, Germany, for funding within the Helmholtz Excellence Network SOLARMATH, a strategic collaboration of the DFG Excellence Cluster MATH+ and Helmholtz-Zentrum Berlin (grant no. ExNet-0042-Phase-2-3). We thank Kerstin Jacob and Matthias Mews (HZB) for substrate preparation. We also acknowledge the support from Alexei Nefedov at the HE-SGM beamline at HZB's synchrotron radiation facility BESSY II. Alexander Hinderhofer, Alexander Gerlach, and Frank Schreiber (University Tübingen) are acknowledged for insightful discussions.